\documentclass{article}

\usepackage{hyperref}
\usepackage{paralist}
\usepackage[UKenglish]{babel}

\title{Regulating Safety and Security in Autonomous Robotic Systems}
\author{Matt Luckcuck and Marie Farrell \\~\\ Department of Computer Science, University of Liverpool, UK}

\begin{document}

\maketitle

Autonomous Robotics Systems are inherently safety-critical and have complex safety issues to consider (for example, a safety failure can lead to a safety failure). Before they are deployed, these systems of have to show evidence that they adhere to a set of regulator-defined rules for safety and security. Formal methods provide robust approaches to proving a system obeys given rules, but formalising (usually natural language) rules can prove difficult. 

Regulations specifically for autonomous systems are still being developed, but the safety rules for a human operator are a good starting point when trying to show that an autonomous system is safe. For applications of autonomous systems like driverless cars and pilotless aircraft, there are clear rules for human operators, which have been formalised and used to prove that an autonomous system obeys some or all of these rules~\cite{Webster2014, rizaldi2017formalising}. However, in the space and nuclear sectors applications are more likely to differ, so a set of general safety principles has developed. This allows novel applications to be assessed for their safety, but are difficult to formalise.

To improve this situation, we are collaborating with regulators and the community in the space and nuclear sectors to develop guidelines for autonomous and robotic systems that are amenable to robust (formal) verification. These activities also have the benefit of bridging the gaps in knowledge within both the space or nuclear communities and academia.

\section{Security Regulations in Space}
\label{sec:space}

On the 1st of February 2019, we held a Space Security Scoping Workshop which was jointly organised by the Universities of Liverpool (Marie Farrell and Michael Fisher) and Warwick (Matthew Bradbury and Carsten Maple). The 29 attendees were from a mix of academia and industry. The aim of this workshop was to discuss the cyber security issues related to robotic systems deployed in space in order to scope out research priorities and to develop collaborative R\&D programmes on the topic of cyber security between FAIR-SPACE and industrial partners.  

Discussions during the workshop revealed that the space industry is becoming more entrepreneurial, with a greater acceptance of risk for more financial gain. Current regulations and standards for space are lacking and often ignored. In particular, rules enforced by the European Space Agency (ESA) when launching satellites can be, and often are, disregarded by other organisations, and so those that do not meet the ESA's requirements may still be launched.

During this workshop, the organisers posed the following five questions to the attendees related to space security.
\begin{compactitem}
\item What are the security issues in space?
\item Are they different to the issues in autonomous ground/air vehicles?
\item What will be the problems in the future?
\item What are current ways of detecting/stopping attacks in these systems?
\item How do environmental considerations impact on security?
\end{compactitem}
From the resulting discussion, it became clear that, although some companies and organisations have a good understanding of the security issues faced by their space systems, there are others that have not seriously considered cyber security. The lack of detailed guidelines and regulations in this area is certainly a stumbling block which makes it difficult for new space companies to know exactly what is required of their systems from a cyber security perspective. There are, however, various guidelines for space systems' security published by the Consultative Committee for Space Data Systems (CCSDS), but it appears that they are no enforced and are not detailed enough, particularly for autonomous robotic missions~\cite{book2006security}.

A full report for this workshop, which describes the discussion of the above questions in detail, has been published by the FAIR-SPACE hub~\cite{space2019}. The report also outlines our future work in this area, which includes several academic publications describing how cyber security threat analysis techniques can be combined with formal verification and some associated case studies. We are in frequent contact with several attendees from the workshop to ensure that our research remains relevant for the space industry. Furthermore, we intend to organise a follow-up workshop in the future, as part of the FAIR-SPACE Hub\footnote{Future AI and Robotics for Space: \url{https://www.fairspacehub.org/}}.

\section{Safety Regulations in the Nuclear Sector}
\label{sec:nuclear}

In the UK, regulation of robotics for nuclear industry is more clear cut. The Office for Nuclear Regulation (ONR) is the government body responsible for checking the safety of any system operating on the `nuclear estate'. While the ONR provide guidance for ensuring system safety, they have not yet produced any guidance specific to autonomous systems. Their guides are also more descriptive than prescriptive, which makes them difficult to formalise and use as a system specification.

To tackle this challenge, we have been running a series of workshops\footnote{Details of the workshops are available at: \url{https://autonomy-and-verification-uol.github.io/events/fnrc}} with the ONR. The workshops aim to be an open forum for discussion between the nuclear operators and supply chain, the ONR, and academia. They explore the safety assessment process for robotic systems in the nuclear industry and examine what may change with the introduction of autonomy. The main focus of the workshops is to clarify (if not answer) the questions surrounding the verification of autonomous robotics. 

We have run two workshops, both attended by a mixture of academics, nuclear operators, robotics developers, and representatives from the ONR. The first workshop introduced the safety assessment process in the UK nuclear industry and verification approaches for autonomous robotic systems, and concluded with a broad discussion session. The second workshop focussed on four case studies of proposed or operational robotic systems from nuclear operators. Two of the case studies were of laser cutting systems, the other two were for remote handling or maintenance. After their introduction, each case study was discussed, in parallel, to examine the hazards and mitigations of the current system and of the same robotic system if it were under autonomous control.


Several issues were raised during the workshops' discussion sessions about the extra considerations needed for the introduction of a robotic system. Firstly, it was thought that a robotic system could widen the environment of the system, to cover the transportation and cleaning or maintenance of the robot, and not just its usual operating facility. There was also recognition that the system failing can bring a human back into the hazardous environment. This points to the need for highly reliable (hardware \textit{and} software) systems, especially if the system is autonomous and especially if it will be used over a long time period. Finally, there were worries about the impact of robotic and autonomous systems on the workforce, both in terms of job availability and lower safety due to complacency. Both of these issues require careful cross-disciplinary study and communication. 

Crucially for the verification of autonomous robotics, the discussions in both workshops revealed that there is no standard good practice for developing robotic or autonomous systems in a way that can be robustly verified. So, to follow up these workshops, we are collaborating with the ONR to develop such guidelines for autonomous robotic systems in hazardous nuclear environments. We are currently drafting the guidelines, which will eventually be opened up to wider consultation. These aim to guide developers to build their systems in ways that make robust (particularly formal) verification easier. These guidelines have the secondary aim of describing the merits and methods of forms of verification that may be new that community.

\bibliographystyle{plain}
\bibliography{papers-regsArticle}

\end{document}